# Low magnetic moment and unconventional magneto-transport in half-Heusler alloy CoVGe

Ravinder Kumar[1], Jyotiraditya Pandey[1], Shoaib Akhtar[1], Sachin Majee[2], Dibyendu Majee[2], Samik DuttaGupta[2] and Sachin Gupta[1,*]

[1]Department of Physics, Bennett University, Greater Noida 201310, India

[2]Saha Institute of Nuclear Physics, A CI of Homi Bhabha National Institute (HBNI), 1/AF, Bidhannagar, Kolkata, West Bengal 700064, India.

*Corresponding author email: sachin.gupta@bennett.edu.in

**Abstract**

In the present work, we experimentally realize CoVGe for the first time and investigate its structural, magnetic, and transport properties, supported by theoretical calculations. The material crystallizes in a cubic structure and exhibits a very low magnetic moment of 0.13 $\mu_B$ per formula unit at 5 K. The temperature dependence of electrical resistivity suggests half-metallic behaviour. Magnetoresistance shows a positive, non-saturating linear field dependence at low temperature that gradually weakens with increasing temperature. The combination of low magnetic moment and unusual magnetotransport behaviour positions CoVGe as a promising platform for exploring spin-dependent transport in Heusler-based materials.

-------------------------------------------------------------------------------------------------

# 1. Introduction

The continuous evolution and miniaturization of modern electronic and spintronic devices have led to an increased sensitivity to intrinsic magnetostatic dipolar interactions.[1–4] As device dimensions shrink, these stray fields become relatively stronger, often resulting in reduced magnetic stability, switching speed, and large energy dissipation.[4,5] Consequently, it is crucial to identify materials that possess intrinsically low magnetostatic field while maintaining high spin polarization. Conventional ferromagnets serve effectively as spin sources; however, they produce large stray fields.[5,6] In contrast, antiferromagnets (AFMs) exhibit fully compensated magnetic sublattices, leading to negligible net magnetization and minimal dipolar fields.[7] Nevertheless, symmetric density of states at Fermi level in conventional AFMs typically yields zero net spin polarization, rendering them unsuitable for direct spin injection or detection. An effective strategy to overcome this limitation involved discovering or engineering materials that simultaneously exhibit vanishing or very low moment and electronically asymmetric spin-dependent density of states.[8,9] Materials that have 100 % spin polarization at the Fermi level with zero net or vanishing magnetic moment are termed as half-metallic antiferromagnets, or more precisely, half-metallic fully compensated ferrimagnets (HMFCFs).[3,8,10–14] Such materials offer high spin polarization with negligible magnetization, offering a promising pathway toward next-generation, energy-efficient spintronic architectures.[5]

Heusler alloys are multifunctional materials which have great potential for spintronic applications, owing to their ability to combine high spin polarization and vanishing magnetic moment. In particular, Heusler materials exhibiting half-metallic fully compensated ferrimagnetism (HMFCF) achieve both features simultaneously—an ideal combination for next-generation spintronic technologies that produces minimal stray fields and efficient spin transport.[8–10,15–17] Although full- and quaternary Heusler can also host HMFCF states, half-Heusler materials are interesting because their structural simplicity facilitates magnetic compensation. In these systems, magnetic compensation arises when transition-metal atoms occupy symmetry-related sublattices whose opposing moments cancel out each other.[16,18] In addition to crystallographic symmetry, the electronic structure of these alloys can be tuned by variations in stoichiometry, enabling precise control over their magnetic and transport characteristics.[17,19] Intrinsic defects—particularly antistites disorder—may further reduce the net moment while preserving high spin polarization.[20–23] Through this combined control of atomic ordering, composition, and defect chemistry, half-Heusler alloys offer a promising and adaptable platform for engineering next-generation spintronic materials.

Previous reports show that CoVGe has 18 valence electrons and follows the Slater–Pauling rule, resulting in a zero total magnetic moment.[24,25] Motivated by these theoretical calculations, we experimentally synthesize CoVGe half Heusler alloy using the arc-melt technique and subsequently investigate its structural, magnetic, and transport properties. The CoVGe crystallizes in a cubic structure. The magnetic properties show very low magnetic moment. The electrical transport measurements suggest half-metallic nature and unconventional magnetoresistance.

## 2. Experimental and computational details

Polycrystalline CoVGe sample was synthesized in an inert argon atmosphere using high-purity constituent elements Co, V, and Ge (≥99.99 at. %). The elements were melted together in a water-cooled copper hearth. To ensure homogeneity, the ingot was remelted 4-5 times by flipping it upside-down. The as-cast sample was then sealed in an evacuated quartz tube and annealed at 850 °C for seven days, followed by ice-water quenching to enhance the degree of crystallinity. The crystal structure of the sample was examined using room temperature powder X-ray diffraction (XRD) performed on a Bruker D8 Advance diffractometer with Cu Kα radiation ($\lambda$ = 1.54 Å). The obtained XRD pattern was analysed through Rietveld refinement using the FullProf Suite software.[26] The magnetic measurements were carried out using a vibrating sample magnetometer (VSM) attached with a physical property measurement system (PPMS, Cryogenic Ltd., UK) as a function of temperature and magnetic field. A rectangular CoVGe pellet (0.83 mm × 1.59 mm × 0.83 mm) was used for transport measurements, with electrical contacts made using high-conductivity silver paste. Temperature-dependent resistivity ($\rho$) was measured in a standard linear four-probe configuration, with current and voltage along the x-direction and the magnetic field applied normal to the sample plane (z-direction) using a PPMS (model 600), Quantum Design, USA. Hall measurements were performed with the current along *x*, the magnetic field along *z*, and the transverse Hall voltage detected along the *y*-direction.

First-principles calculations were performed using the Vienna ab initio Simulation Package (VASP) with the projector augmented wave (PAW) method.[27–29] The generalized gradient approximation (GGA) was employed for the exchange-correlation energy, as parametrized by Perdew, Burke, and Ernzerhof (PBE).[30] A 15 × 15 × 15 k-point mesh and the experimental lattice parameters obtained from X-ray diffraction were used for the density of states (DOS) calculations.

## 3. Results and Discussions

### 3.1 Crystal structure

Half Heusler alloys are represented by an empirical formula; *XYZ* (where, *X* and *Y* are transition metals and *Z* is a *p*-block element) and crystallize in a non-centrosymmetric cubic structure (space group no. 216, $F\bar{4}3m$, $Cl_b$).[31] The Half-Heusler structure consists of three interpenetrating face-centered cubic (fcc) sublattices, with the *X*, *Y*, and *Z* atoms occupying distinct lattice sites.[13,31] The atoms occupy the Wyckoff positions 4*a* (0, 0, 0), 4*b* (½, ½, ½), and 4*c* (¼, ¼, ¼), respectively.[13,31] Fig. 1. shows the room-temperature powder X-ray diffraction (XRD) pattern along with the Rietveld refinement. The XRD pattern exhibits (111) and (200) superlattice peaks, along with the (220) principal peak. Additionally, a few un-indexed impurity peaks, marked with asterisks (*), can also be seen. The presence of superlattice peaks confirms ordered structure in CoVGe. The lattice constant obtained from the Rietveld refinement is *a* = 5.88 Å.

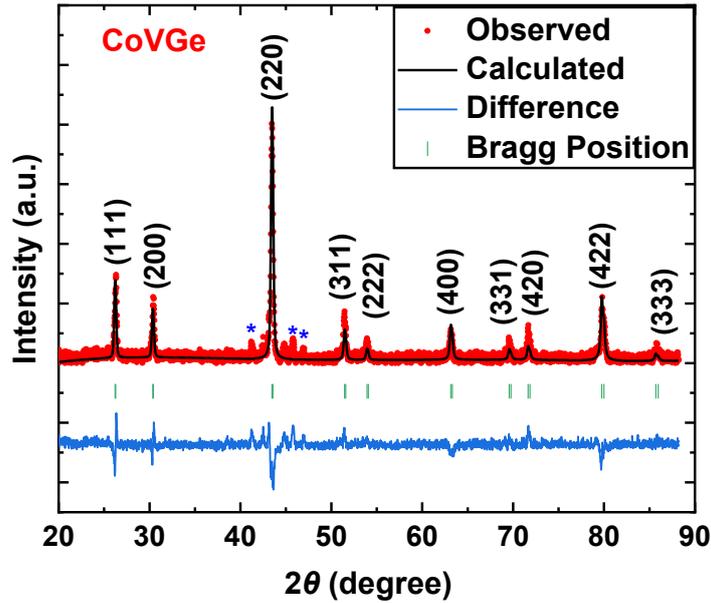

**FIG. 1.** Room temperature powder X-ray diffraction pattern of CoVGe along with the Rietveld refinement.

### 3.2 Magnetic properties

Fig. 2(a) presents the zero-field-cooled warming (ZFCW) and field-cooled warming (FCW) thermo-magnetic (*M–T*) curves of CoVGe, measured under an applied magnetic field, *H* of 500 Oe. The Curie temperature, $T_c$ of CoVGe, estimated from the first derivative of the *M–T* curve, is approximately 43 K, as illustrated in the inset of Fig. 2(a). *M-T* plots show small thermomagnetic irreversibility at low temperatures which might attributed to domain wall pinning effect.[32,33] It is worth noting that there is no signature of additional magnetic transitions, ruling out significant magnetic contribution from impurity, if present. Fig. 2(b) presents the magnetization, *M* as a function of magnetic field, *H* at various temperatures (5, 20, 40, 100, and 300 K). The *M–H* curves reveal that CoVGe exhibits soft magnetic ordering below 43 K, as evidenced by the characteristic S-shaped narrow non-zero hysteresis behaviour. Above this temperature, the CoVGe has paramagnetic nature, demonstrated by a linear *M–H* relationship in Fig. 2(b). The magnetic moment measured at 5 K is approximately 0.13 $\mu_B$/f.u. and close to the value predicted from Slater–Pauling rule (0 $\mu_B$/f.u.) for this alloy.[23]

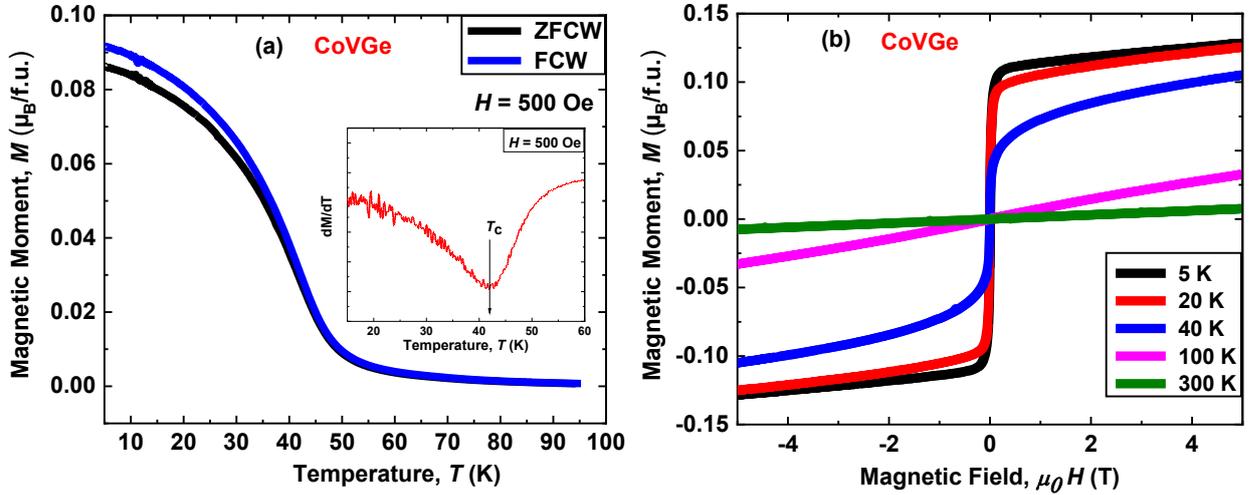

**FIG. 2.** (a) The temperature, *T* dependence of magnetization, *M* curves for zero-field-cooled warming (ZFCW) and field-cooled warming (FCW) at *H* = 500 Oe. (b) The field, *H* dependence of magnetization, *M* as a function of temperature, *T*.

### 3.3 Transport properties
### 3.3.1. Electrical resistivity

The temperature dependence of resistivity, *ρ*(T) was measured at zero magnetic field in the range of 2–300 K for CoVGe as shown in Fig. 3. The electrical resistivity increases with temperature, exhibiting a metal-like behaviour. The residual resistivity ratio (RRR = $\rho_{300K}$ / $\rho_{2K}$) was found to be 1.6. To examine the half-metallic nature of the material, the electrical resistivity data were fitted in the temperature range of 2–43 K using the power-law equation, expressed as:[34,35]

$$\rho(T) = \rho_o + AT^n \tag{1}$$

Where $\rho_o$ represents the residual resistivity of the material, $A$ is an arbitrary constant and $n$ is an exponent. The solid red line in Fig. 3 shows the fitting of resistivity data using equation (1) in the range of 2-43 K. In conventional ferromagnets, the resistivity typically exhibits a quadratic ($T^2$) temperature dependence arising from electron–magnon scattering.[36–38] However, in half-metallic systems, this $T^2$ contribution is expected to vanish because single-magnon scattering is prohibited due to presence of an energy gap at Fermi level in the minority-spin channel.[39] An exponent $n = 1.4$ obtained from fit using equation (1) indicates a deviation from the typical quadratic trend, implying that electron–magnon scattering is significantly suppressed owing to the absence of minority-spin carriers at low temperatures. This indicates the half metallic nature of the material.[40]

At higher temperatures (43-300 K), the temperature-dependent resistivity was fitted using a model that incorporates the combined scattering contributions, described as:[40]

$$\rho(T) = \rho_o + \rho_{ph} + \rho_{mag} \qquad (2)$$

Where, $\rho_{ph}$ denotes the electron–phonon scattering term, while $\rho_{mag}$ corresponds to the magnonic scattering contribution. The temperature-dependency of these scattering can be expressed as:

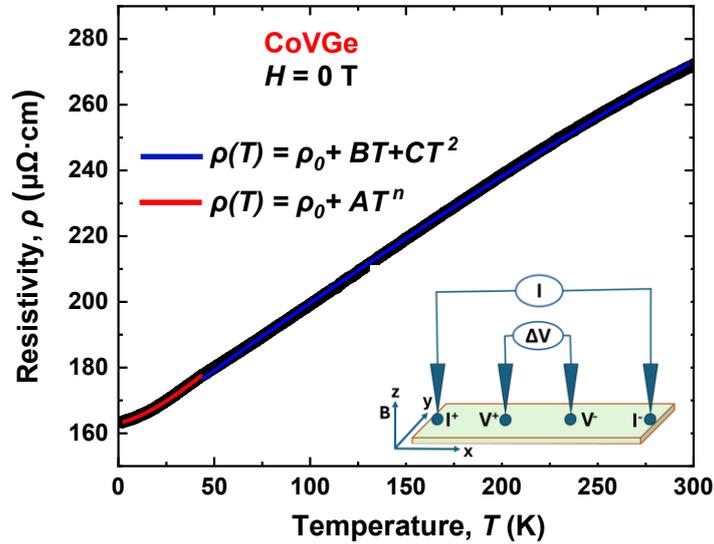

**FIG. 3.** The temperature dependence of electrical resistivity, $\rho$ for CoVGe, measured in the temperature range 2-300 K. The solid red and blue lines are fit to experimental data using equations (1) & (3), respectively. Inset shows the measurement geometry.

$$\rho(T) = \rho_o + BT + CT^2 \qquad (3)$$

Where, *B* and *C* are constants associated with phonon and magnon scattering, respectively. Equation (3) fits the experimental data well, as shown by the solid blue line in Fig. 3, and the extracted parameters are listed in Table 1. The value of *B* suggests that electron- phonon scattering is dominant at higher temperatures.

**Table 1:** List of parameters obtained from low and high temperatures electrical resistivity fit using equations (1) and (3), respectively.

| Material | Low temperature fitting parameters | | | | High temperature fitting parameters | | | |
|---|---|---|---|---|---|---|---|---|
| | $\rho_o$ ($\mu\Omega$.cm) | A ($\mu\Omega$.cm K$^{-n}$) | n | Range (K) | $\rho_o$ ($\mu\Omega$.cm) | B ($\mu\Omega$.cm K$^{-1}$) | C ($\mu\Omega$.cm K$^{-2}$) | Range (K) |
| CoVGe | 163.2 | 0.07288 | 1.4 | 2 - 43 | 157.7 | 0.43611 | 1.69 ×10$^{-4}$ | 43-300 |

### 3.3.2. Hall resistivity

Hall effect measurements were conducted to further investigate the electronic behaviour of the material. The Hall resistivity, $\rho_H$ was measured at various temperatures under an applied magnetic field, *H* up to ±5 T as shown in Fig. 4. A clear anomalous Hall effect was observed only at 5 K as anticipated for ferromagnetic materials. At higher temperatures (100, 200, and 300 K), the material shows the ordinary Hall effect due to its paramagnetic nature. In ferromagnetic systems, the Hall effect typically arises from the combined contributions of the ordinary Hall effect and the anomalous Hall effect.
The total Hall resistivity $\rho_H$ can be written as:[41,42]

$$\rho_H = R_o \mu_o H + R_s M \qquad (4)$$

The first and second term in the above equation represent the ordinary Hall and anomalous Hall contribution. Here, $R_o$ corresponds to the ordinary Hall coefficient, and $R_s$ to the anomalous Hall coefficient, whereas $\mu_0$ is the magnetic permeability of free space.[41] *M* is saturation magnetization of the material. The ordinary Hall coefficient $R_o$ was used to estimate the carrier concentration at 5 K using equation, $n = 1/(R_0 e)$. The value of $R_o$ was obtained from the linear fit of the high-field Hall resistivity data, yielding a carrier concentration of the order of $10^{22}$ cm$^{-3}$, which is comparable with other reported half metallic materials.[43] The Anomalous Hall coefficient which is intercept of Hall resistivity plot at 5 K is coming out to be 0.46 μΩ-cm. In the high-temperature regime (100-300K), above the Curie temperature, Hall resistivity exhibits almost linear behavior anticipated for paramagnetic materials.

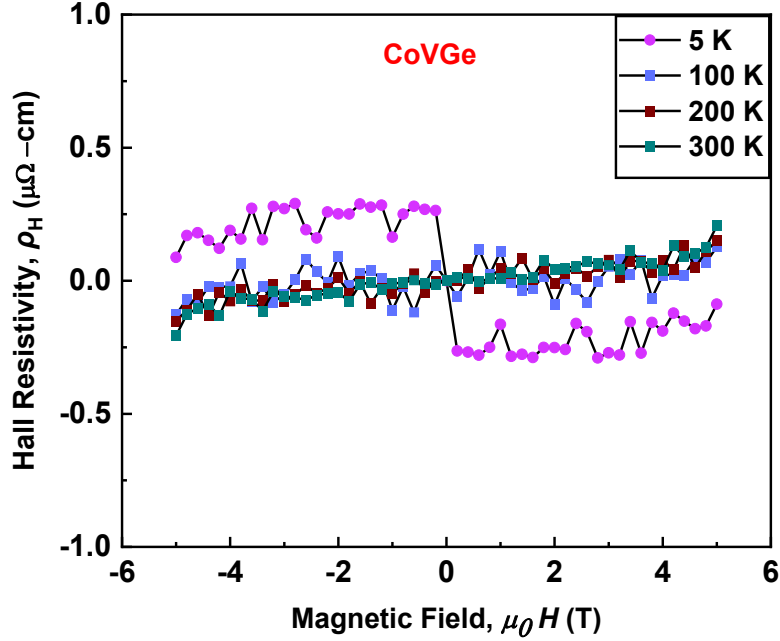

**FIG. 4.** The field dependence of Hall resistivity, $\rho_H$ measured at different temperatures (5, 100, 200, and 300 K).

### 3.3.3. Magnetoresistance (MR)

Fig. 5 presents the magnetic-field dependence of magnetoresistance (MR) in CoVGe measured at different temperatures. At 5 K, CoVGe exhibits a clear non-saturating linear positive magnetoresistance (LPMR) up to ±5 T, as guided by a linear fit to the MR–H data. In contrast to conventional ferromagnetic metals, where negative MR is typically observed due to the suppression of spin-dependent scattering with increasing magnetic field, the positive MR observed here cannot be explained by classical Lorentz-force-driven orbital magnetoresistance, which anticipates a quadratic field dependence (MR ∝ H²) and is not observed in the present case.[44,45]

The observed linear positive magnetoresistance (LPMR) at low temperature is more plausibly associated with the intrinsic electronic structure of CoVGe near the Fermi level. Similar linear MR behaviour has been reported in several Heusler and related compounds, such as $Fe_2CoSi$, $Mn_2CoAl$, CoRuVSi, and CoFeVSi, where it has been attributed to gapless or nearly gapless electronic states, spin-gapless semiconducting behaviour, or band crossings close to the Fermi energy.[46,45,47,45] In such systems, linear MR is commonly linked to unconventional charge transport arising from peculiar band topology or strong band-structure effects rather than classical orbital motion.

With increasing temperature, the MR decreases and becomes negligible above 100 K, indicating that thermal scattering and phonon-induced broadening suppress the mechanisms responsible for LPMR. At 300 K, the weak negative MR observed is consistent with conventional spin-disorder scattering in ferromagnetic materials.

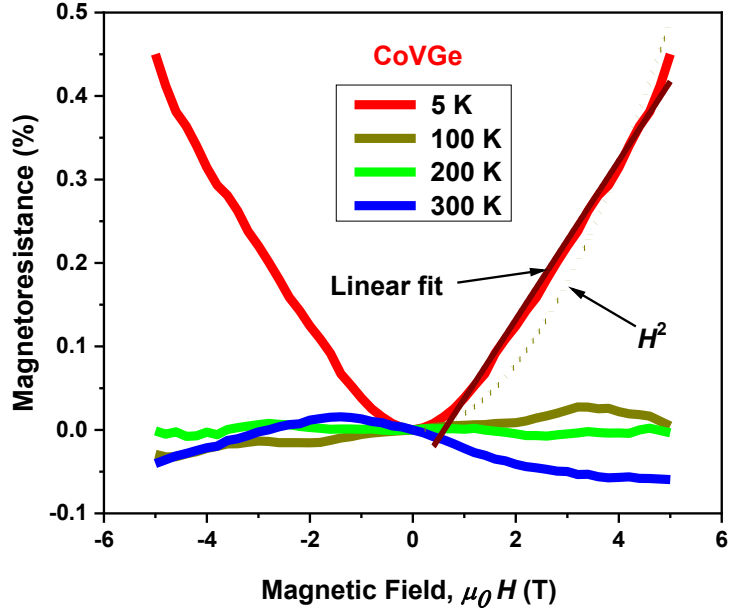

**FIG. 5.** The field dependence of magnetoresistance, MR for CoVGe at different temperatures.

### 4. Theoretical results

Fig. 6 shows the density of states (DOS) as a function of energy, $E$ for CoVGe, calculated using first principles calculations. The calculations are performed by taking the lattice constant, $a = 5.88$ Å for ordered crystal structure, observed experimentally from XRD data. The DOS for the majority (spin-up) and minority (spin-down) channels are represented by red and blue colours, respectively. In the majority spin channel, density of states (DOS) is present at the Fermi level, indicating metallic conduction behaviour. The minority spin channel also exhibits a small finite DOS at the Fermi level, although a suppression of states is observed slightly below the Fermi energy, forming a pseudogap-like feature. For an ideal half-metallic material, one spin channel should exhibit metallic behaviour while the other shows a semiconducting or insulating gap at the Fermi level.[48] However, in CoVGe, the minority spin channel does not display a complete gap at the Fermi energy; instead, it exhibits a pseudogap-like feature with a finite density of states at the Fermi level. These results differ from the theoretically predicted semiconducting behavior reported by L. Mohan *et al.* using a relaxed lattice constant (6.30 Å), suggesting that the electronic structure of CoVGe is sensitive to lattice parameters and structural details.[25]

The calculated density of states of CoVGe indicates the presence of a small finite states and a pseudogap-like feature near the Fermi energy in the minority spin channel, which may be associated with band crossings or a near-gapless electronic structure. Such features could contribute to the experimentally observed linear positive magnetoresistance at low temperatures. In Heusler compounds, linear positive magnetoresistance has frequently been linked to gapless or nearly gapless electronic structures, spin-gapless semiconducting behaviour, or band crossings in the vicinity of the Fermi level.[46, 45,47,45]

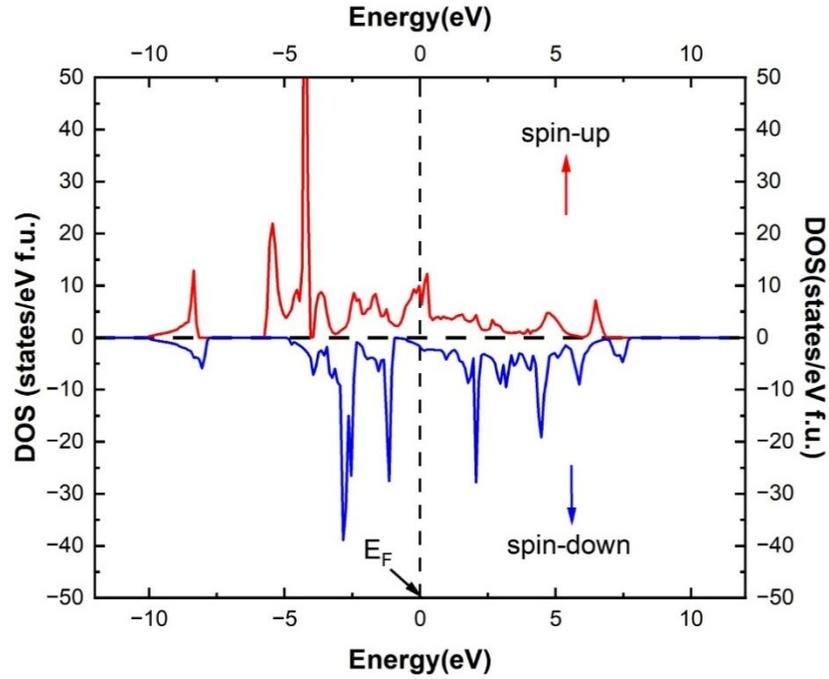

**FIG. 6.** Density of states (DOS) of CoVGe. The majority spin channel (spin-up) and minority spin channel (spin-down) are represented by red and blue colours, respectively.

Conclusions

Half Heusler alloy CoVGe was synthesized using arc melt technique, which crystallizes in cubic structure. Magnetic measurements show soft-ferromagnetic nature, and a very low magnetic moment of 0.13 $\mu_B$/f.u. at 5 K. Electrical resistivity measurements suggest half-metallic behaviour. The linear positive magnetoresistance observed at 5 K was possibly due to small gap or gapless electronic structure near the Fermi level. A very low value of magnetic moment and unconventional MR in CoVGe could make this material interesting for exploring spin-dependent transport in Heusler-based materials.

## Acknowledgements

The authors sincerely thank Prof. K. G. Suresh, IIT Bombay, for providing the material synthesis facilities. S.G. acknowledges the financial support from the Anusandhan National Research Foundation (ANRF), New Delhi, through project SUR/2022/004713, for the successful completion of this work. S.D.G. acknowledges financial support from the Anusandhan National Research Foundation (ANRF), New Delhi, under project no. ANRF/ECRG/2024/004649/PMS.